\documentclass[epj]{webofc}
\usepackage[varg]{txfonts}   
\woctitle{International Conference on New Frontiers in Physics 2013}
\begin{document}
\title{V+jets production at the CMS}
%
%

\author{Bugra Bilin\inst{1}\fnsep\thanks{\email{bugra.bilin@cern.ch}}, on behalf of the CMS collaboration
}

\institute{Middle East Technical University 
          }

\abstract{%
Measurements of Vector Boson production in association with jets are presented, using p-p collision data at $\sqrt{s} = 7$ TeV. The measurements presented include Z + jets azimuthal correlations, event shapes, vector boson + jets differential cross section measurements, hard double-parton scattering using W + jets events and electroweak Z + forward - backward jet production.
}
\maketitle
\section{Introduction}
\label{intro}

Vector boson production is one of the best understood processes at hadron colliders. Events of W and Z bosons decaying into electrons and muons  are among the cleanest experimental final states. Measurements of jet production associated with vector bosons provide a stringent test of perturbative QCD (pQCD). These processes  are sensitive to new physics effects, and are backgrounds for new physics and Higgs boson searches. These measurements can also provide constraints on Parton Distribution Functions (PDFs) and  are used to improve the Monte Carlo generators. In this paper, various V + jets measurements carried out by the CMS experiment~\cite{CMS} are summarised. 

\section{Z + jets azimuthal correlations and event shape measurement }
\label{sec_1}


Measurement of azimuthal correlations and event shapes of Z + jets process is carried out by the CMS experiment \cite{ref:event_shapes}. The measurement is made in two phase space regions; inclusive (transverse momentum, $P_T(Z)>0$ GeV) and boosted regimes ($P_T(Z)>150$ GeV). Boosted regime is of particular interest since it is important for new physics searches.

\begin{figure}
\begin{minipage}[b]{0.45\linewidth}
\centering
\includegraphics[width=7cm,clip]{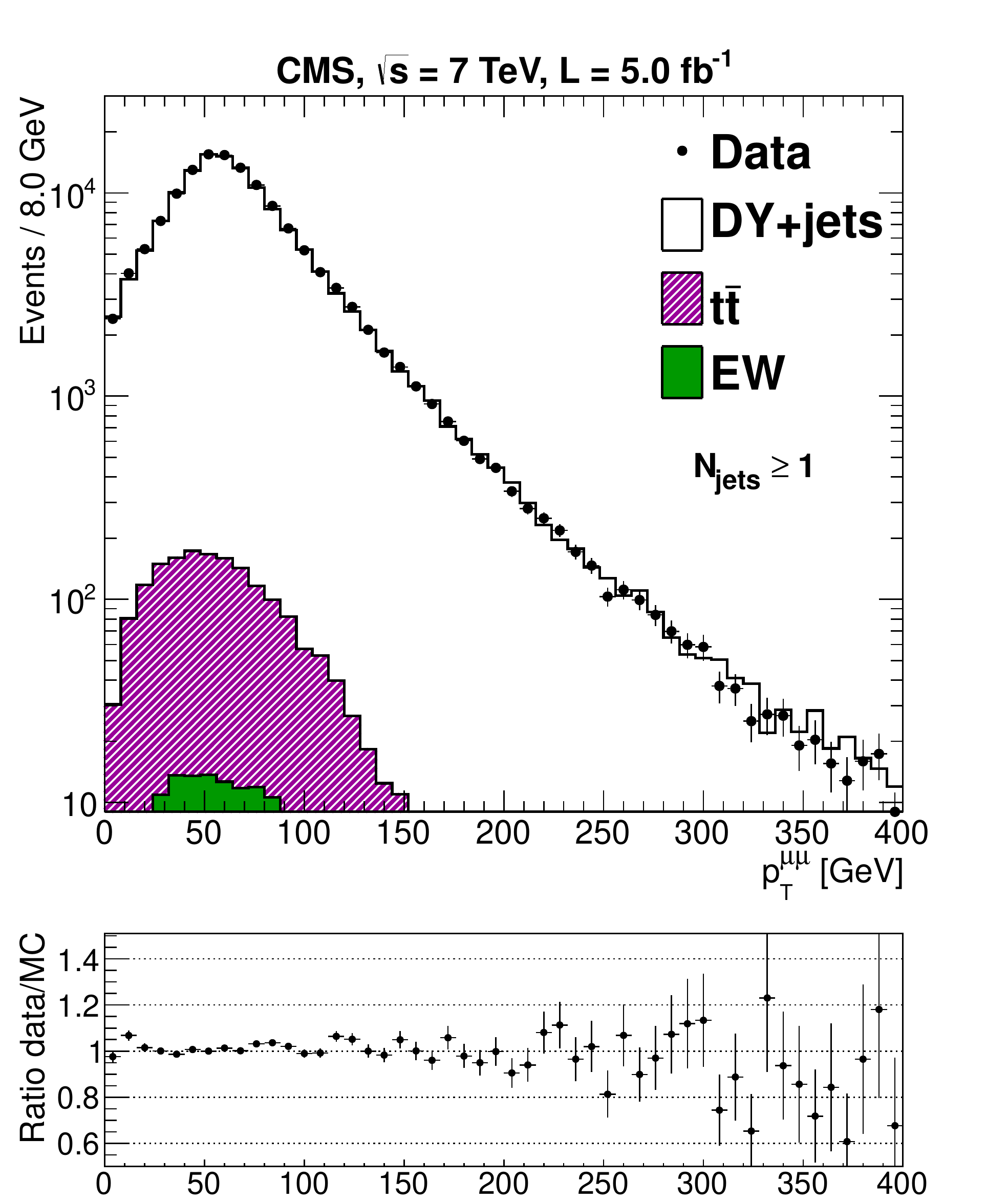}
\caption{Detector level di-muon transverse momentum, after detector efficiency corrections applied.}
\label{fig-1.1}       
\end{minipage}
\hspace{0.5cm}
\begin{minipage}[b]{0.45\linewidth}
\includegraphics[width=7cm,clip]{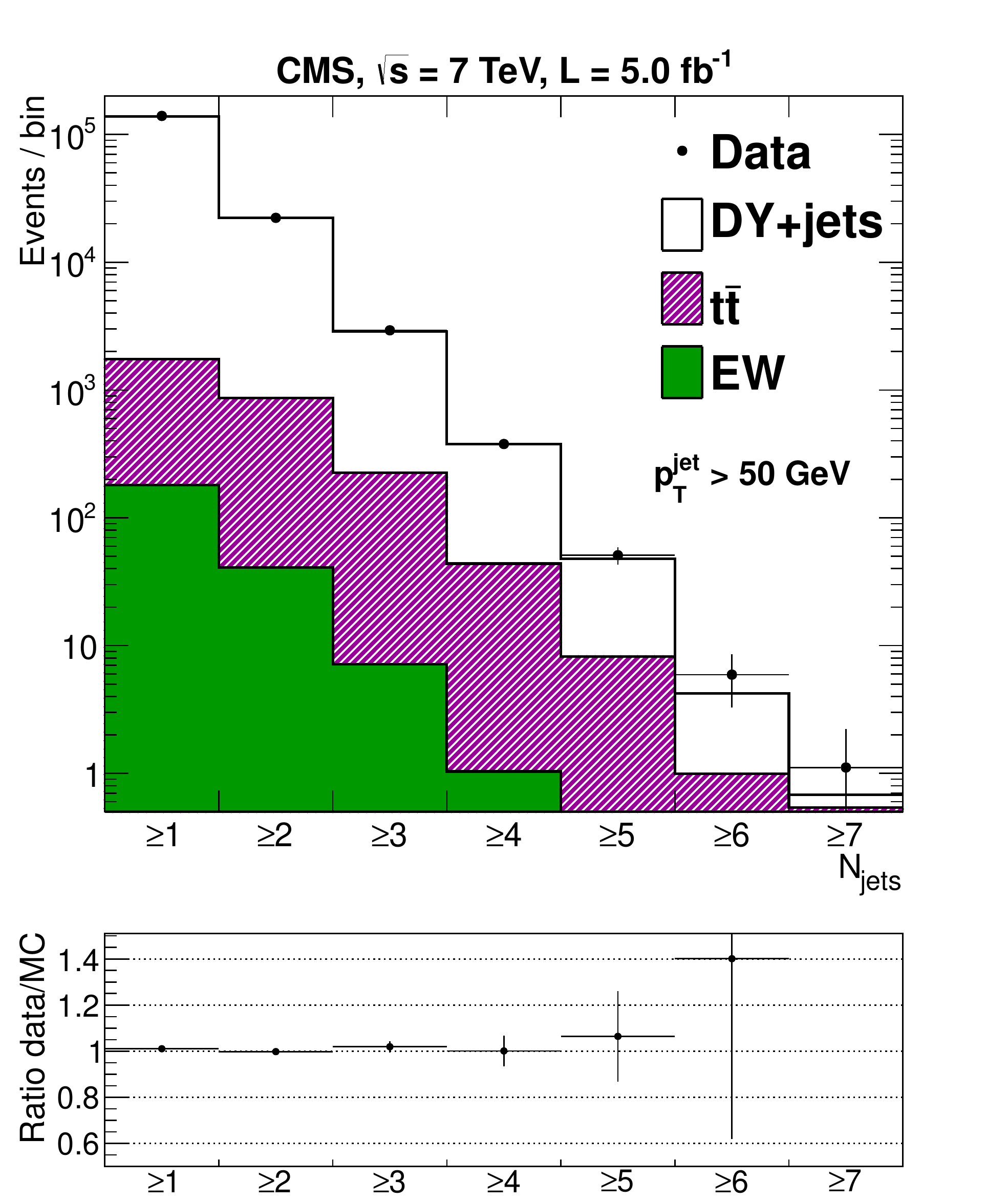}
\caption{Detector level exclusive jet multiplicity for the di-muon events, after detector efficiency corrections applied.}
\label{fig-1.2}       
\end{minipage}
\end{figure}

\begin{figure}
\begin{minipage}[b]{0.45\linewidth}
\centering
\includegraphics[width=7cm,clip]{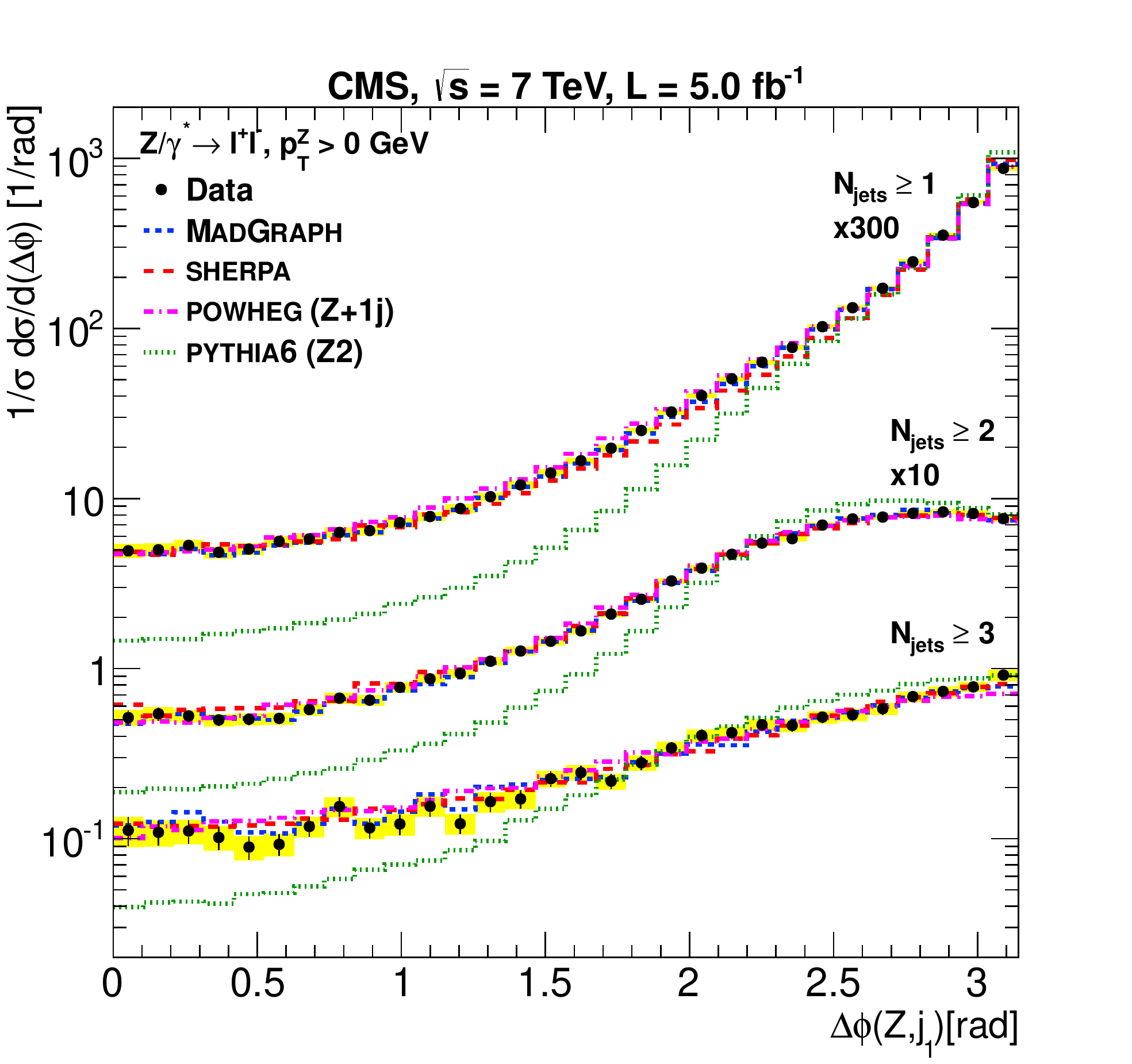}
\caption{Azimuthal angle difference between the Z boson and the leading jet in the inclusive 3 jet multiplicity $\Delta\phi(Z,j_1)$ without any requirements on $P_T(Z)$}
\label{fig-1.3}       
\end{minipage}
\hspace{0.5cm}
\begin{minipage}[b]{0.45\linewidth}
\includegraphics[width=7cm,clip]{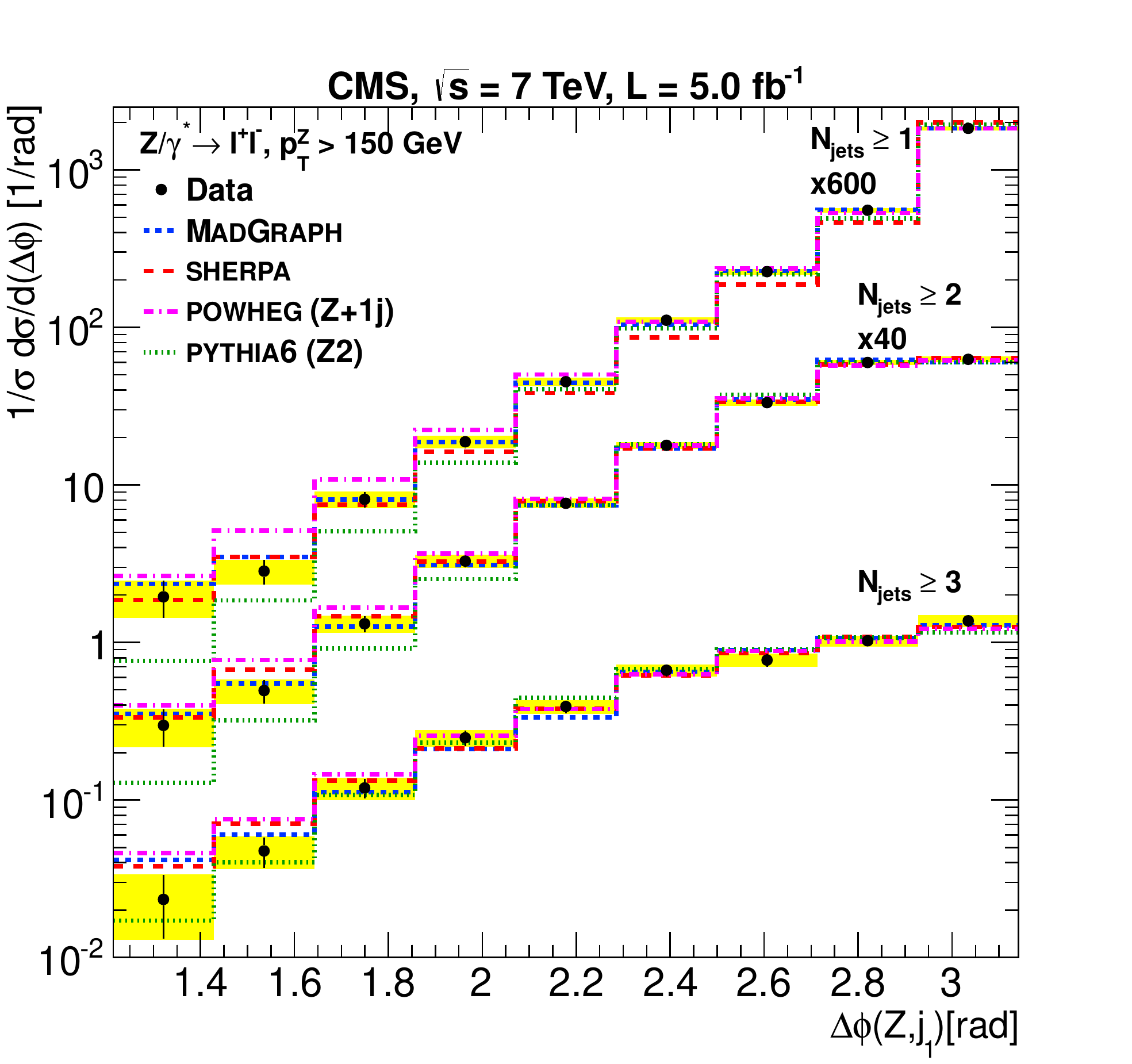}
\caption{Azimuthal angle difference between the Z boson and the leading jet in the inclusive 3 jet multiplicity $\Delta\phi(Z,j_1)$ for the boosted ($P_T(Z)>150.$ GeV) regime}
\label{fig-1.4} 
\end{minipage}
\end{figure}

Figure~\ref{fig-1.1} and~\ref{fig-1.2}  show the $P_T$ distribution of di-muon candidates and the jet multiplicity associated with these events respectively. In Figures~\ref{fig-1.3} and~\ref{fig-1.4}, the azimuthal angle difference between the leading jet and the Z boson are presented for inclusive and boosted regimes respectively. It is seen that the data is well described by the Monte Carlo.

\section{Measurement of Z + 1 jet and photon + 1 jet rapidity distributions}
\label{sec_2}
\begin{figure}
\begin{minipage}[b]{0.45\linewidth}
\centering
\includegraphics[width=7cm,clip]{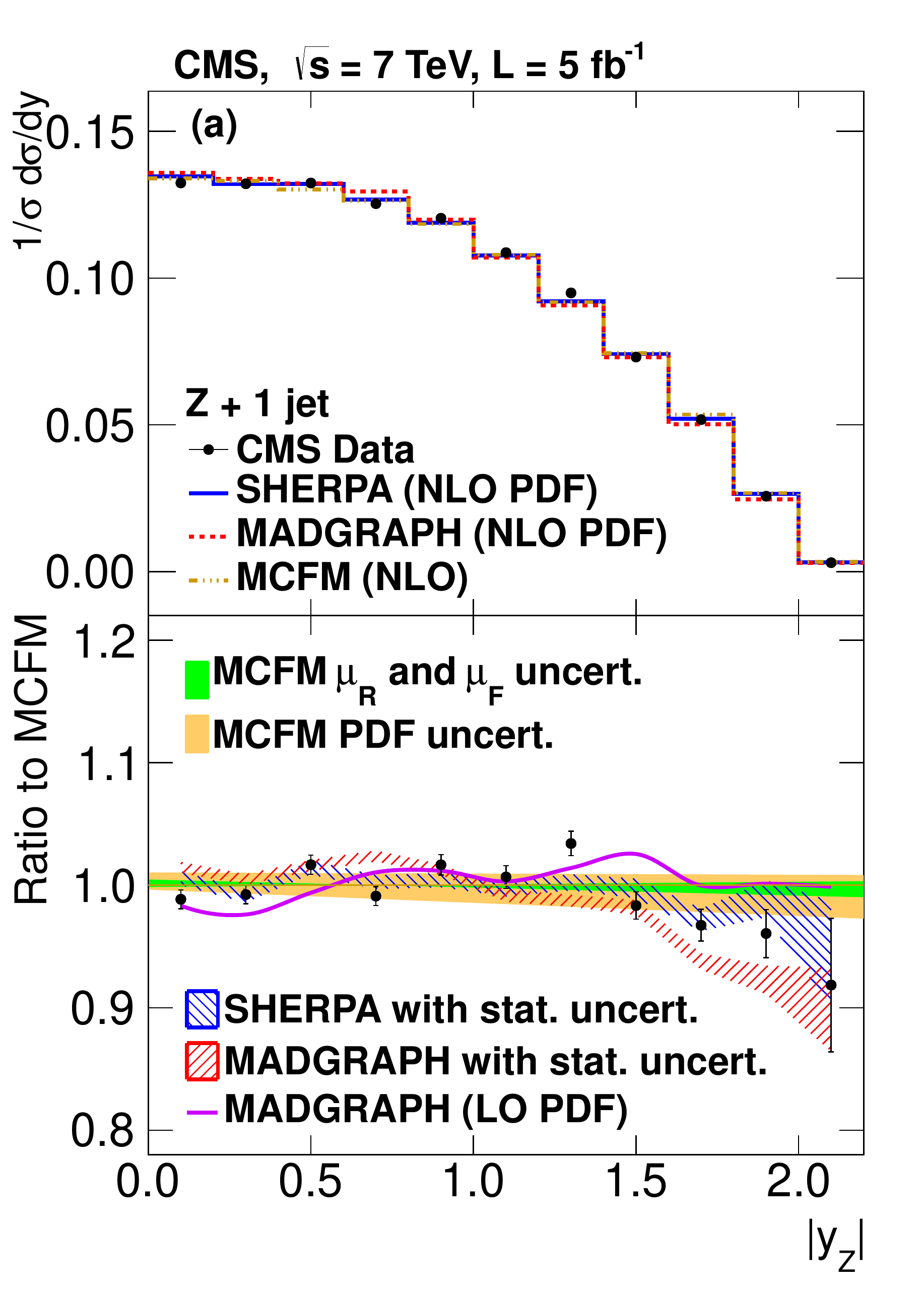}
\caption{Absolute value of Z boson rapidity, normalised to the total cross section. Data is shown after efficiency corrections. Statistical and systematic uncertainties are added in quadrature.}
\label{fig-3.1}       
\end{minipage}
\hspace{0.5cm}
\begin{minipage}[b]{0.45\linewidth}
\includegraphics[width=7cm,clip]{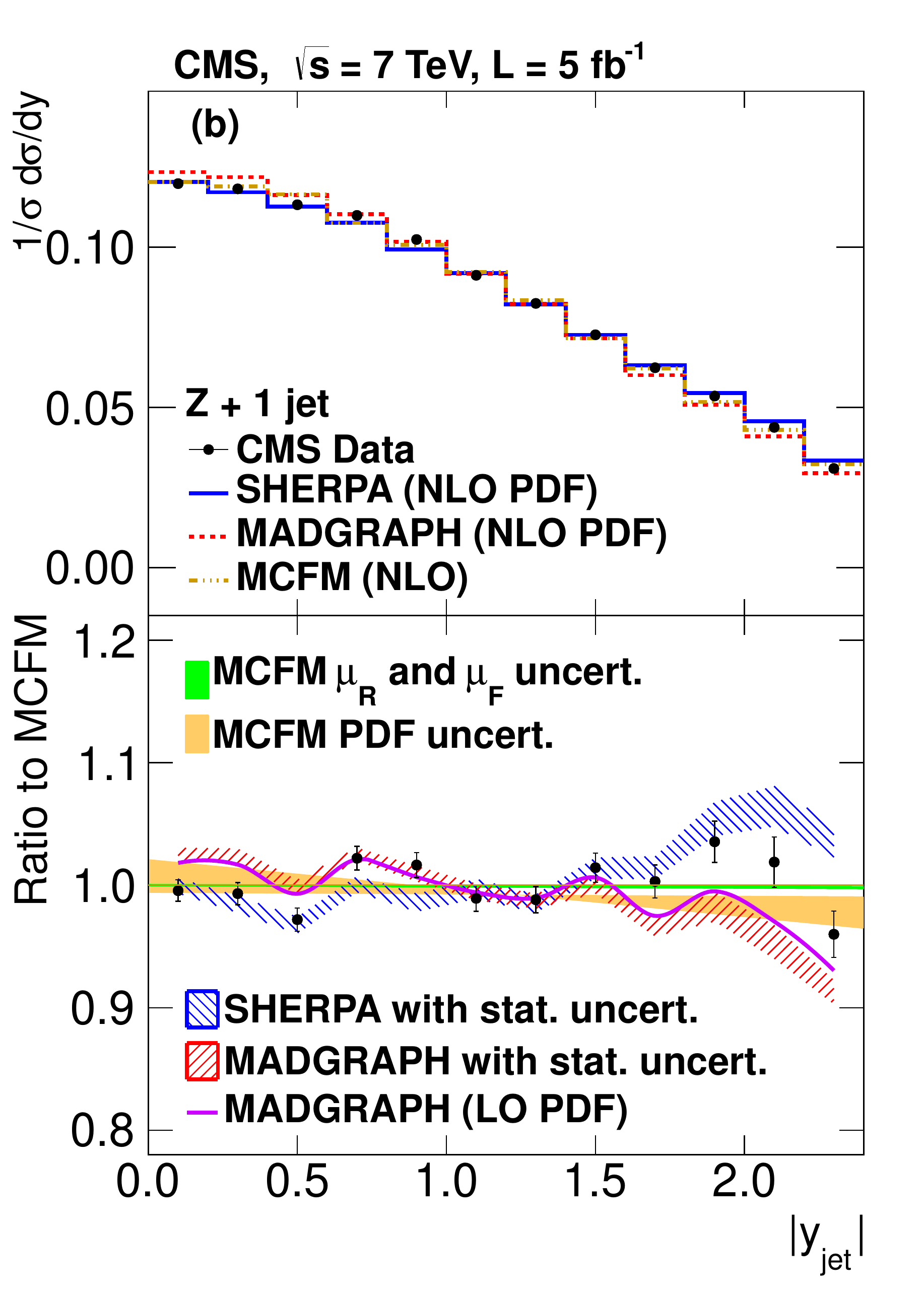}
\caption{Absolute value of the jet rapidity, normalised to the total cross section. Data is shown after efficiency corrections. Statistical and systematic uncertainties added in quadrature.}
\label{fig-3.2}       
\end{minipage}
\end{figure}

Presence of the electroweak vertex makes the perturbative QCD calculations stable. Precise measurement of the rapidity in Z+jet events is required for Higgs boson measurements. Measurement of rapidities in events with a vector boson ($\gamma$, Z) and a jet is performed  \cite{ref:rapidity}.
The measurement of vector boson and the jet rapidities, the rapidity difference ($y_{dif}=|y_V - y_{jet}|/2$), and the rapidity sum  ($y_{sum}=|y_V + y_{jet}|/2$) is carried out.   Here $y_{sum}$ is related to the boost from laboratory frame to the centre of mass frame of the V+jet system. The quantities $y_{sum}$ and  $y_{diff}$ are uncorrelated observables, whereas $y_Z$ and $y_{jet}$ are highly correlated. The rapidity sum, $y_{sum}$ is mainly dependent on the PDFs, whereas $y_{diff}$ is sensitive to Leading Order (LO) partonic differential cross sections. Figures \ref{fig-3.1} - \ref{fig-3.2} show $y_Z$ and $y_{jet}$ distributions for Z + jet events.

Vector boson and jet rapidity distributions are shown to be consistent with the perturbative QCD predictions, while $y_{sum}$  and $y_{diff}$ distributions show discrepancies between QCD predictions and data. The calculations at the Next to Leading Order (NLO) QCD, and simulations using two different Matrix Element to Parton Shower matching methods fail to describe data and are inconsistent among each other.  

\section{Photon+jets differential cross section measurement}
\label{sec_3}
Measurements of photon production with one or more jets ($\gamma$ + jets) provide a stringent test of  pQCD. The production cross sections of $\gamma$ + jets for various angular configurations are strongly dependent on QCD hard-scattering sub-processes. These processes are also directly sensitive to the gluon PDFs in the proton, hence may provide constraints on the PDFs. These measurements provide tests of fixed order pQCD calculations in a wide range of kinematic phase space.

\begin{figure}
\begin{minipage}[b]{0.45\linewidth}
\centering
\includegraphics[width=7cm,clip]{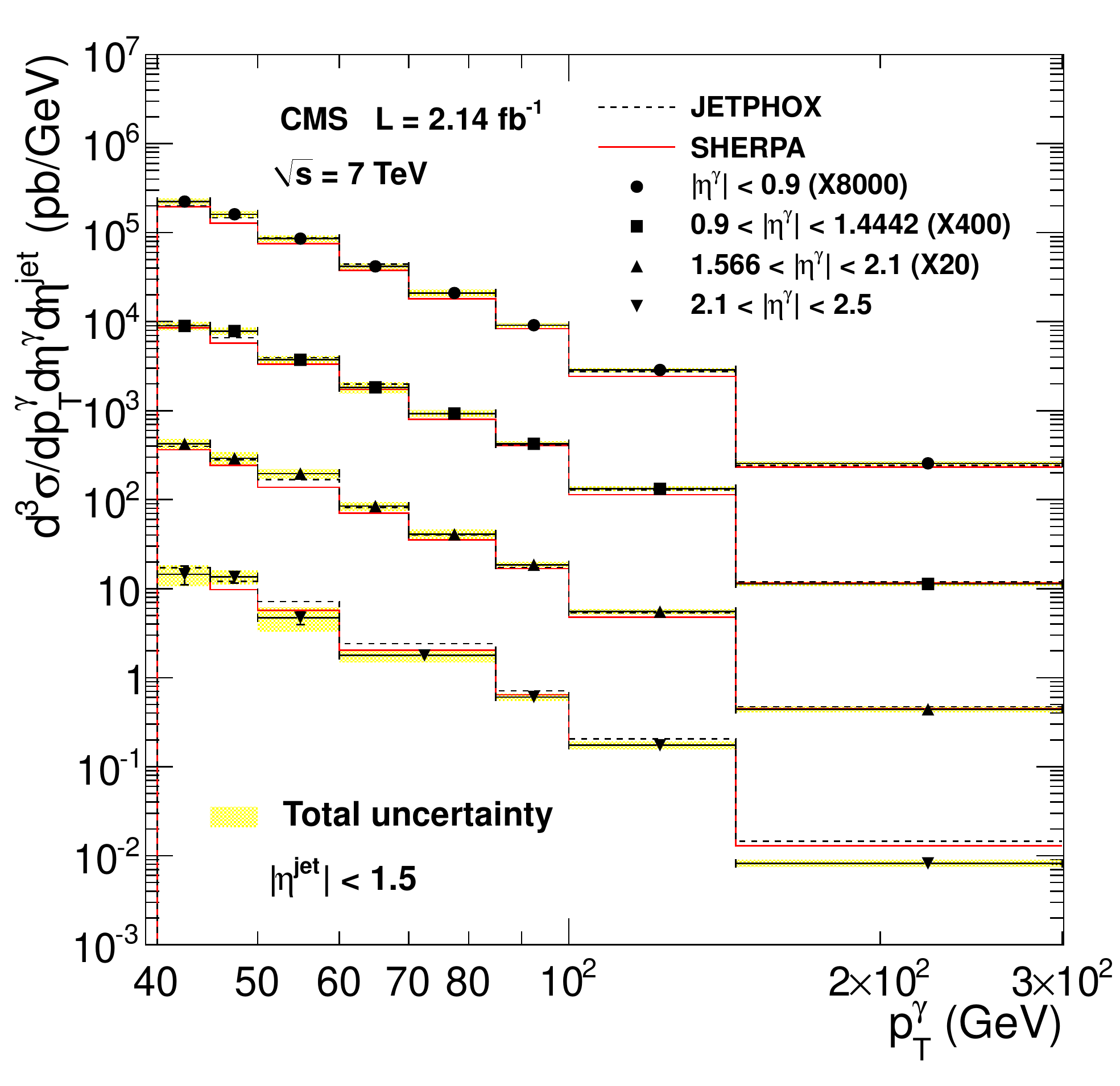}
\caption{Cross-sections for $|\eta|<1.5$. Error bars are statistical uncertainties and yellow bands are the total uncertainties obtained by adding in quadrature statistical and systematic uncertainties.}
\label{fig-2.1}       
\end{minipage}
\hspace{0.5cm}
\begin{minipage}[b]{0.45\linewidth}
\includegraphics[width=7cm,clip]{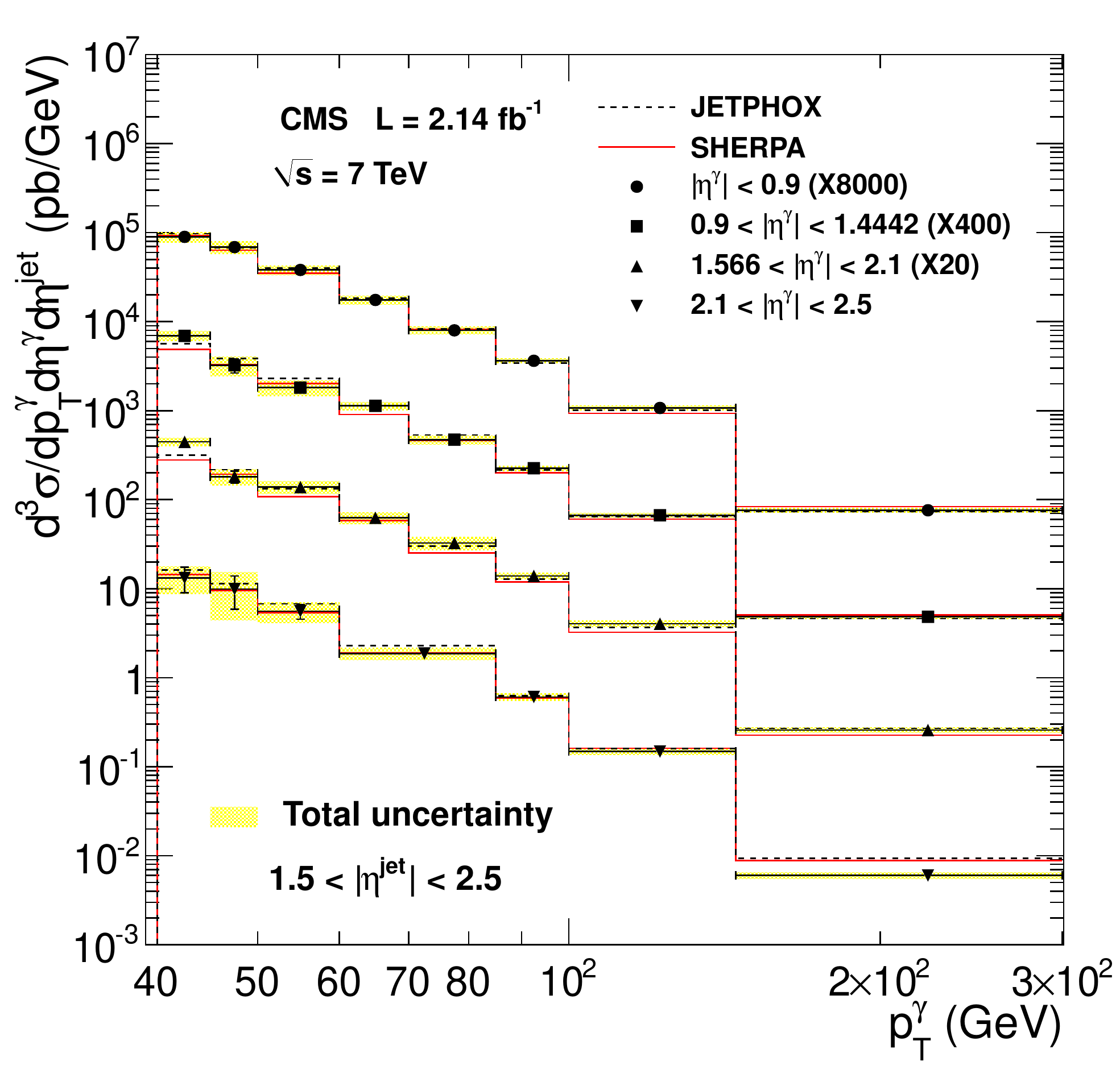}
\caption{Cross-sections for 1.5<$|\eta|<2.5$. Error bars are statistical uncertainties and yellow bands are the total uncertainties obtained by adding in quadrature statistical and systematic uncertainties.}
\label{fig-2.2}      
\end{minipage}
\end{figure}

Triple differential cross section of $\gamma$ + jet processes, $d^3\sigma/dp_T^{\gamma} d\eta^{\gamma}d\eta^{jet}$,  is measured \cite{ref:gamma_jet} as well as ratios of the cross sections with respect to different jet-photon orientations are measured, where some of the main uncertainties cancel. The cross section distributions are displayed in Figures~\ref{fig-2.1} and \ref{fig-2.2}.
The results show that the NLO predictions describes the data well, whereas LO predictions underestimate the data.

\section{Di-jet mass spectrum measurement in W + 2 jets}
\label{sec_4}

CDF Collaboration reported an excess in invariant mass ($M_{JJ}$) constructed from the two highest transverse momentum jets in events with W boson and two jets in the mass range 120 - 160 GeV \cite{ref:cdf_wjet}. The cross section of excess is determined to be $4$ pb. A similar search is carried out by the D0 collaboration and an upper limit of 1.9 pb is set on the cross section at 95 \% CL \cite{ref:d0_wjet}. An updated analysis is reported by the CDF collaboration \cite{ref:cdf_wjet2} using full statistics available. With improved calibrations of the detector response and a better understanding of the instrumental backgrounds, the standard measurements were found to be in agreement with the standard model monte carlo predictions. 

\begin{figure}
\centering
\includegraphics[width=7cm,clip]{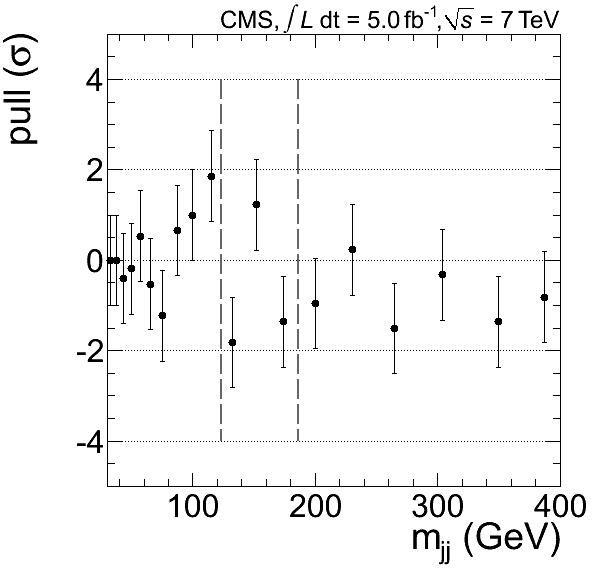}
\caption{Pull distribution ([data - fit] / fit uncertainty) vs. $M_{jj}$. The dashed line represent the searched signal region which is excluded from the fits.}
\label{fig-4.3}       
\end{figure}

Motivated by the former "CDF bump", search for a di-jet resonance is carried out  in W+jet events by CMS for W + 2 jet and 3 jet final states \cite{ref:cms_wjet}. 
The contributions of standard model processes to di-jet invariant mass is determined using an unbinned maximum-likelihood fit in the invariant mass range 40 - 400 GeV, excluding the signal region, 123 - 186 GeV. The difference of data from the fit, divided by the fit uncertainty (pull) is shown in Figure \ref{fig-4.3}. No excess is observed, and an upper limit of 5.0 pb is set at 95\% CL on the production cross section for a generic Gaussian signal with a mass about 150 GeV.

\section{Measurement of double parton scattering in W+jets}
\label{sec_5}

Small momentum fractions (x) carried by the incoming partons are accesible via proton-proton collisions at the LHC energies. The large number of partons having small momentum  fractions increase the probability of two parton-parton scattering occuring at the same time and producing 2 identifiable hard scattering in proton-proton interactions. Double Parton Scattering (DPS) studies might provide information on spatial structure of hadrons, and provide improved understanding of backgrounds to new physics and higgs boson searches. 

DPS is investigated in events with leptonically decaying W boson and 2 jet final states \cite{ref:DPS}. DPS events differ from the Single Parton Scattering (SPS) events due to the fact that the di-jet system from one parton interaction is independent from the W coming from the other parton interaction. In this study, variables sensitive to discriminate DPS from SPS events; the azimuthal angle difference between jets ($\Delta \phi(j1,j2)$), the ratio of the magnitude of vectorial sum of the $p_T$ of jets to the scalar sum of $p_T$ of jets ($\Delta^{rel} p_T $), and the azimuthal angle between W and dijet system  ($\Delta S$) are measured. 

\begin{figure}
\begin{minipage}[b]{0.45\linewidth}
\centering
\includegraphics[width=7cm,clip]{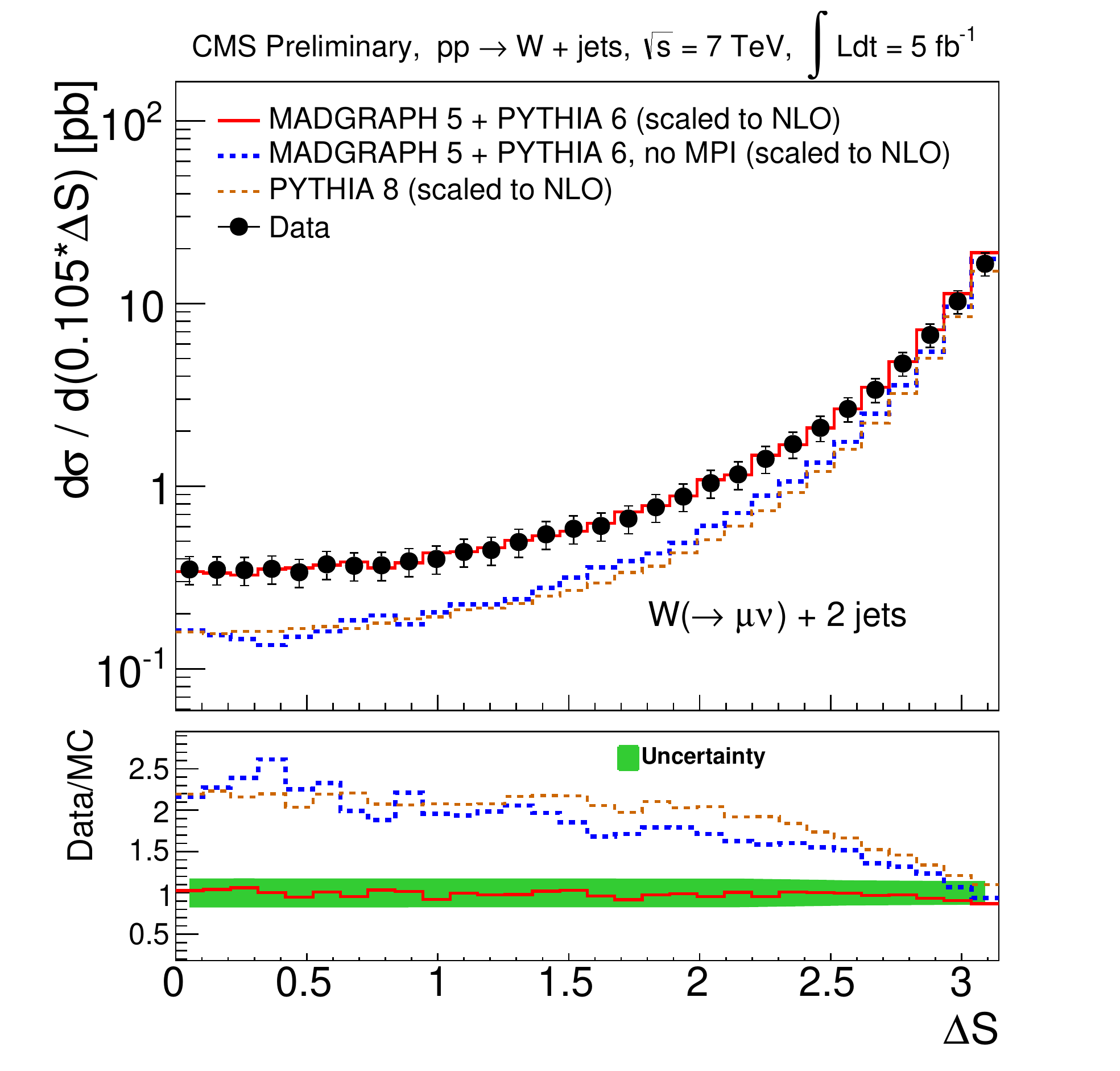}
\caption{Fully-corrected differential cross sections for $\Delta S$. }
\label{fig-5.1}       
\end{minipage}
\hspace{0.5cm}
\begin{minipage}[b]{0.45\linewidth}
\includegraphics[width=7cm,clip]{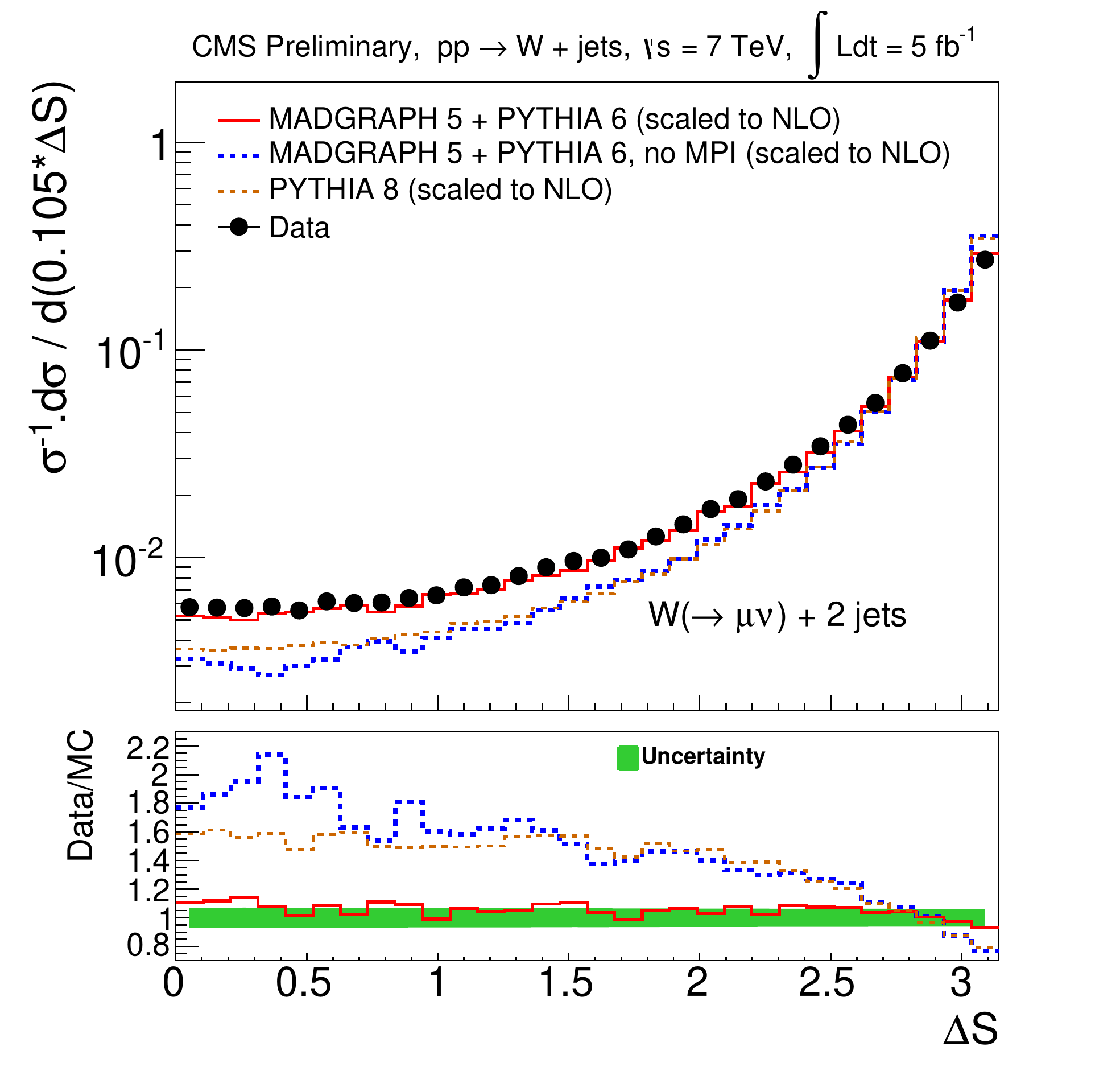}
\caption{Fully-corrected differential cross sections for $\Delta S$, normalised to the total cross section.}
\label{fig-5.2}       
\end{minipage}
\end{figure}

In Figures~\ref{fig-5.1} and~\ref{fig-5.2}, fully corrected differential cross sections are shown for the $\Delta S$ variable for events with exactly two jets accompanying the W boson. Monte Carlo predictions including DPS simulation describe the data well, whereas the simulations without DPS fails to describe the cross section and the shape of the distribution.
This study is the first step towards extracting the underlying DPS fraction at LHC energies. 

\section{Electroweak Z + forward-backward jets production measurement}
\label{sec_6}

Pure electroweak production of W and Z bosons with two well separated jets is quite sizeable at the LHC \cite{ref:EWKZ2J}. Study of these process is important especially for Higgs production from  Vector Boson Fusion. Measurements of the pure electroweak production of Z boson with two forward backward jets is carried out in $ll jj$ final state.
To extract the small electroweak signal from the Drell Yan background, two different methods have been carried out. Firstly, the signal is extracted from a fit to the $m_{j_1j_2}$ distribution. As a cross check, a multivariate analysis technique is used, where a boosted decision tree with decorrelation (BDTD) is trained using various observables. The signal is hence confirmed by two different approaches.
The cross section is measured as $\sigma=154\pm24$(stat.) $\pm46$ (exp. syst.) $\pm27$ (th. syst.)  $\pm3$ (lum.) fb, which is consistent with the theoretical prediction at NLO, $166$ fb.

\section{Conclusions}
\label{sec_7}

Measurements of various V+jet processes with CMS detector are presented in this paper. These measurements provide a detailed description of  the topological structure of V+ jets events. The measurements test the validity of QCD, and provide confidence for existing Monte Carlo models in describing standard model processes.

 Some other important V+jet measurements are carried out by CMS experiment which are not covered in the talk, such as Heavy Flavour production with Vector bosons (V+HF) \cite{ref:Zb, ref:Wbb, ref:Wc, ref:Zbb} that provide constraints to the heavy quark PDFs.   

In all measurements, good agreement is observed between data and standard model Monte Carlo predictions.

\end{document}